\documentclass[12pt]{article}

\usepackage[utf8]{inputenc}
\usepackage[T1]{fontenc}
\usepackage{times}
\usepackage[margin=1in]{geometry}
\usepackage{graphicx}
\usepackage{booktabs}
\usepackage{array}
\usepackage{tabularx}
\usepackage{hyperref}
\usepackage{url}
\usepackage{natbib}
\usepackage{setspace}
\usepackage{titlesec}
\usepackage{parskip}
\usepackage{textcomp}
\usepackage{hanging}

\hypersetup{
    colorlinks=true,
    linkcolor=black,
    citecolor=black,
    urlcolor=blue,
    bookmarks=true
}

\titleformat{\section}{\large\bfseries}{\thesection.}{0.5em}{}
\titleformat{\subsection}{\normalsize\bfseries}{\thesubsection.}{0.5em}{}

\begin{document}

\begin{center}
{\Large\bfseries Developing the PsyCogMetrics\texttrademark{} AI Lab to Evaluate Large Language Models
and Advance Cognitive Science\\---A Three-Cycle Action Design Science Study}

\vspace{1.5em}

\begin{tabular}{ccc}
Zhiye Jin & Yibai Li & K.\ D.\ Joshi \\
Marywood University & The University of Scranton & University of North Carolina Wilmington \\
\href{mailto:zjin@m.marywood.edu}{zjin@m.marywood.edu} &
\href{mailto:yibai.li@scranton.edu}{yibai.li@scranton.edu} &
\href{mailto:joshik@uncw.edu}{joshik@uncw.edu} \\[1em]
Xuefei (Nancy) Deng & Xiaobing (Emily) Li & \\
California State University & The University of Scranton & \\
Dominguez Hills & & \\
\href{mailto:ndeng@csudh.edu}{ndeng@csudh.edu} &
\href{mailto:xiaobing.li@scranton.edu}{xiaobing.li@scranton.edu} & \\
\end{tabular}
\end{center}

\vspace{1.5em}

\noindent\textbf{Abstract}

\noindent
This study presents the development of the PsyCogMetrics\texttrademark{} AI Lab (\url{https://psycogmetrics.ai}), an integrated, cloud-based platform that operationalizes psychometric and cognitive-science methodologies for Large Language Model (LLM) evaluation. Framed as a three-cycle Action Design Science study, the Relevance Cycle identifies key limitations in current evaluation methods and unfulfilled stakeholder needs. The Rigor Cycle draws on kernel theories such as Popperian falsifiability, Classical Test Theory, and Cognitive Load Theory to derive deductive design objectives. The Design Cycle operationalizes these objectives through nested Build--Intervene--Evaluate loops. The study contributes a novel IT artifact, a validated design for LLM evaluation, benefiting research at the intersection of AI, psychology, cognitive science, and the social and behavioral sciences.

\vspace{0.5em}

\section{Introduction}

Evaluation is fundamental to the development of large language models \citep{Chang2023}. Evaluation tracks progress and establishes benchmarks that guide future improvements. Many training methods such as reinforcement learning depend on a quantified reward function (i.e., evaluation), without which there is nothing to optimize and thus no path to improvement. However, despite the proliferation of benchmarks and metrics, current evaluation methods for large language models (LLMs) still suffer from benchmark saturation, data contamination, and lack of coverage. The challenge stems from the inherent complexity of LLMs, which parallels that of the human brain. Psychologists, cognitive scientists, and social and behavioral scientists have a long history of studying the even more complex human brain and behavior. They have instruments and methodologies, but current LLM evaluation tools are overwhelmingly developer oriented. The lack of accessible, easy-to-use, and integrated platforms limits their ability to contribute their expertise to LLM evaluation.

Motivated by this gap, the present study introduces PsyCogMetrics\texttrademark{} AI Lab, a novel design-science artifact developed to bridge the fields of psychology, cognitive science, social and behavioral science, and AI model evaluation. Rooted in the cognitivist perspective on artificial intelligence (AI), the platform applies psychometric, cognitive science, and experimental social-behavioral methodologies to investigate the internal structures, behavioral tendencies, and reasoning abilities of LLMs. It is a fully operational, cloud-native system, co-designed with domain stakeholders and refined through iterative action design research cycles. The PsyCogMetrics\texttrademark{} AI Lab, available at \url{https://psycogmetrics.ai}, aims to democratize access to rigorous LLM evaluation by integrating visual modeling tools, reproducible workflows, and scientifically grounded metrics into a unified, user-friendly interface.

This study has two key objectives. First, it seeks to develop and evaluate an IT artifact---the PsyCogMetrics\texttrademark{} AI Lab---that meets stakeholder-defined needs for robust, rigorous, explainable, easy to use, and integrated LLM evaluation. Second, it aims to advance design theory by integrating kernel theories from the philosophy of science, Classical Test Theory and Cognitive Load Theory into each phase of artifact development, thereby providing guidance for future development of similar LLM evaluation artifacts.

Building on \citeauthor{Hevner2007}'s \citeyearpar{Hevner2007} three-cycle method of design science research, the rest of the paper is organized as follows. Section~2 presents the Relevance Cycle that outlines the real-world problem space, stakeholder requirements, and gaps in existing tools, all of which justify the need for the PsyCogMetrics\texttrademark{} AI Lab. Then, Section~3 discusses the Rigor Cycle, grounding the study in established theoretical frameworks that inform the design decisions. Following that, Section~4 describes the Design Cycle, detailing the objectives, methodology, implementation, and iterative build--intervene--evaluate (BIE) loops used to construct and refine the artifact. Section~5, the discussion section, reflects on the study's contributions to design science research in light of \citet{Hevner2004}'s seven guidelines, highlighting novelty, relevance, rigor, and effective communication.

\section{Relevance Cycle}

Guided by the design science premise that an artifact must both emerge from and deliver value to its problem environment \citep{Hevner2007}, the Relevance Cycle plays a critical role in grounding the research in real-world needs. It involves (1)~identifying the problem space and relevant stakeholders, (2)~scanning the field for related work, and (3)~recognizing gaps in existing solutions. In this study, the Relevance Cycle channels contextual requirements---derived from the psychometric and cognitive evaluation of large language model (LLM) practices---into the design of the PsyCogMetrics\texttrademark{} AI Lab \citep{Hevner2004}.

\subsection{Problem Space}

The problem space of this study lies within the domain of LLM evaluation, a task widely acknowledged as challenging due to their inherent complexity. These models have intricate architectures with vast parameters, making them opaque black boxes. Even the developers of these models often struggle to interpret why they exhibit certain behaviors \citep{LiPsychometrics2025}.

While model developers often focus on the engineering and operational aspects of LLM evaluation---such as computational, memory, energy, financial, and network efficiency across various lifecycle stages including architecture design, pretraining, fine-tuning, and system deployment \citep{Bai2024}---broader stakeholders such as end users, organizations, and regulatory bodies are primarily concerned with the cognitive evaluation of LLMs. Such evaluation typically falls into three key categories: capability (knowledge and reasoning), alignment (bias, toxicity, truthfulness), and safety (robustness and potential misuse) \citep{Ouyang2022, Yi2023}.

Regarding evaluation approaches, current methodologies generally fall along a spectrum between two philosophical perspectives on AI: instrumentalism and cognitivism. Instrumentalism views AI primarily as a practical tool rather than a genuine thinking entity: its value lies in its utility, not in replicating human cognition \citep{Dennett1987, Dennett1991}. From this standpoint, AI systems are ``instruments'' that transform inputs into outputs without requiring consciousness or genuine understanding.

Under instrumentalism, LLM evaluation methods focus on syntax level performance metrics and task-specific benchmarks as pragmatic proxies for utility, emphasizing mechanistic explanations rather than evidence of genuine understanding. Automatic metrics such as perplexity \citep{DataCamp2024}, BLEU \citep{Papineni2002}, and ROUGE \citep{Lin2004} measure token-level predictive quality or n-gram overlap against human references, treating lower perplexity or higher overlap scores as indicators of an LLM's efficacy without implying any cognitive comprehension. Human evaluation frameworks, which gather ratings on fluency, coherence, preference, or alignment, are widely used to guide model development through techniques like Reinforcement Learning from Human Feedback (RLHF) \citep{Ouyang2022}. Emerging reference-free techniques \citep{Yi2023} use either free-form natural-language feedback from humans or auxiliary LLMs as ``AI graders'' to assess outputs without relying on ground-truth references, which further exemplify this tool-centric stance.

In contrast, cognitivism holds that artificial systems can genuinely replicate human thought by treating mental states as computational states \citep{Fodor1975, Putnam1967}. According to the Computational Theory of Mind (CTM)---first articulated by Putnam and later refined by Fodor---cognition is the manipulation of symbolic representations through algorithms executed on neural ``hardware,'' whether biological or artificial \citep{McCulloch1943}. This view is aligned with the Strong AI position, defended by \citet{Hofstadter1995} and \citet{Kurzweil2005}, which asserts that a program of sufficient complexity not only mimics human behavior but also achieves true understanding and consciousness.

Building on the cognitivist view that LLMs can be probed much like human minds, GLUE aggregates performance across nine classic NLU tasks---such as sentiment analysis (SST-2), inference (MNLI), and question-answer entailment (QNLI)---to produce a single score reflecting generalization over diverse linguistic phenomena, treating the model as if it possesses human-like semantic and inferential skills \citep{Wang2018}. SuperGLUE introduces more demanding challenges---coreference resolution (Winograd-style WSC), commonsense reasoning (COPA), and multi-sentence inference (RTE)---alongside a diagnostic suite for pinpointing specific reasoning shortcomings, thereby aligning even more closely with human cognitive abilities \citep{Wang2019}. Measuring Massive Multitask Language Understanding (MMLU) evaluates an LLM's accuracy on 57 disparate multiple-choice subjects drawn from high-school and college-level exams, ranging from world history and government to mathematics and ethics \citep{Hendrycks2021}. The MMLU framework treats the LLM as a learner with a broad ``mental'' knowledge base. Scoring well on MMLU is taken as evidence that the model approximates facets of human expertise across many disciplines \citep{Hendrycks2021}.

The PsyCogMetrics\texttrademark{} AI Lab adopts a cognitivist approach, aiming to evaluate large language models specifically based on psychometric and cognitive science methodologies \citep{LiPsychometrics2025}. Researchers have applied this approach in prior studies: \citet{LiBlockchain2024} adapt standard self-report instruments (e.g., the Big Five) and compare LLM ``self-assessments'' with behavioral measures; COBBLER \citep{Liu2024} designs bias-inducing prompts to quantify anchoring, framing, confirmation, and other classic cognitive biases; Theory of Mind vignettes \citep{Strachan2024, WangToM2025} test LLMs' capacity to attribute beliefs and intentions, highlighting both human-like success and LLM-specific reasoning patterns; \citet{Strachan2024} show GPT-4 reaching six-year-old human parity on several ToM vignettes. Psychometric adaptive testing \citep{Zhu2023, Trismik2024} employs Item Response Theory to efficiently estimate models' latent abilities with far fewer questions. In a recent study, \citet{LiPsychometrics2025} further validated this approach by demonstrating that models such as GPT-4 and LLaMA-3 not only satisfy core psychometric validity criteria---including convergent, discriminant, predictive, and external validity---but also outperform earlier models like GPT-3.5 and LLaMA-2 across these dimensions.

\subsection{Related Work}

The PsyCogMetrics\texttrademark{} AI Lab aims to serve as an IT artifact for LLM evaluation. This section specifically discusses the existing IT artifacts in this field, which mainly include code libraries (CLI/API) and community-driven evaluation arenas.

One of the code libraries, the lm-eval-harness provides a unified command-line interface (CLI) and Python API that supports zero-shot and few-shot benchmarking across a wide range of tasks \citep{EleutherAI2023}. OpenAI Evals simplifies the process of creating and running custom evaluations, blending automated metrics with human review, and includes a companion cookbook for practical implementation \citep{OpenAI2023}. LLMPerf extends the evaluation by incorporating performance metrics such as latency and cost alongside accuracy \citep{RayProject2023}. Meanwhile, Ragas focuses on retrieval-augmented generation (RAG) pipelines, evaluating outputs based on faithfulness, recall of provided context, and hallucination detection \citep{Kulkarni2024}.

Community-driven evaluation platforms, also known as arenas or leaderboards, represent a distinct IT artifact within the LLM evaluation ecosystem. These platforms differentiate themselves by enabling dynamic, peer-driven assessments through large-scale aggregation of real-world human judgments and community-submitted tasks. Rather than relying solely on static benchmarks, they use pairwise model comparisons, voting mechanisms, and crowd-sourced feedback to create live, evolving leaderboards \citep{Chiang2024, LMSYS2023}. Platforms like Chatbot Arena and Hugging Face Leaderboards exemplify this approach \citep{Sethuraman2024, HuggingFace}. Additionally, initiatives such as Search Arena and Wikibench extend this paradigm into retrieval-augmented evaluation and collaborative dataset curation \citep{Kuo2024}. By integrating elements of all traditional categories yet relying fundamentally on open human input, these platforms advance transparency and adaptivity in LLM evaluation \citep{Bhavsar2025}.

\subsection{Gaps}

Despite the proliferation of benchmarks and metrics, current methods for LLM evaluation do not adequately meet the varied demands of key stakeholders. For AI developers and researchers, popular benchmarks quickly become saturated---new models routinely achieve near-ceiling scores without delivering real capability improvements (the benchmark saturation problem) \citep{Koehn2025}---and static test sets often leak into training corpora, artificially inflating evaluation results (the data contamination problem). At the same time, as LLMs evolve, many benchmarks simply cannot capture emerging capabilities (the lack of coverage problem) \citep{McIntosh2024}.

End users, regulators and policymakers demand transparent, interpretable outputs to foster trust and usability, robust safety, fairness, and compliance but evaluation methods fall short in this regard \citep{Bhatt2019}.

LLMs have become attractive subjects and tools in experimental psychology, cognitive science and social scientists, because they can simulate reasoning tasks and generate large samples at negligible cost \citep{Niu2024}. However, lack of accessible, easy to use and integrated platforms limit them contributing their expertise to LLM evaluation. Existing toolkits and libraries are overwhelmingly developer-oriented: they presume strong programming skills, deep familiarity with LLM frameworks, and skills of setting up the software and hardware infrastructure.

These problems and unaddressed stakeholder needs create significant gaps, motivating the development of the PsyCogMetrics\texttrademark{} AI Lab to fill them.

\section{Rigor Cycle}

The Rigor Cycle positions PsyCogMetrics\texttrademark{} AI Lab squarely within the established knowledge base, ensuring that each design decision is traceable to---and advances---cumulative theory. Guided by the three-cycle view, the cycle ``pulls'' justificatory knowledge from the literature before each build and ``pushes'' new prescriptive insights back to that literature \citep{Hevner2007}.

The philosophy of science, Classical Test Theory and Cognitive Load Theory are kernel theories that provide foundational design principles guiding the subsequent stages of the Design Cycle of PsyCogMetrics\texttrademark{} AI Lab.

The PsyCogMetrics\texttrademark{} AI Lab aspires to be a rigorous platform for scientific research. Accordingly, we ground our design in established theories from the philosophy of science that articulate the essential requirements for scientific tools. One key framework informing our approach draws from the perspectives of Karl Popper \citep{Popper2005} and Thomas Kuhn \citep{Kuhn1962}.

Karl Popper famously maintained that scientific theories cannot be definitively verified, only falsified \citep{Popper2005}. In this view, reproducible observations are essential for falsification: ``non-reproducible single occurrences are of no significance to science'' unless they can be independently repeated \citep{Popper2005}. Only when an observation is reproducible does it carry the epistemic weight required to challenge a universal hypothesis \citep{Popper2005}. Thus, reproducibility is not a mere methodological check but the logical basis by which potential refutations acquire scientific legitimacy \citep{Popper2005, SEP2018}.

Reproducibility can be distinguished from related concepts such as repeatability and replicability. Reproducibility, in its narrow technical sense, refers to re-evaluating the same dataset with the same analysis pipeline to confirm results. Repeatability denotes the ability of the same researcher or team to obtain consistent results using identical materials and methods, often in a single laboratory. Replication, by contrast, involves independently conducting a new experiment or observational study to see if similar conclusions emerge under similar conditions \citep{SEP2018}. Within the philosophy of science, these distinctions serve different epistemic functions: reproducibility bolsters statistical and analytical soundness; replication assesses the robustness and generalizability of findings; and repeatability confirms operational reliability of methods \citep{SEP2018}.

To meet the scientific rigor expected by researchers, the design of the PsyCogMetrics\texttrademark{} AI Lab is grounded in Classical Test Theory (CTT) and psychometric validity principles. Classical test theory posits that an observed score ($X$) is composed of a true score ($T$) and a random error component ($E$), such that $X = T + E$ \citep{Lord1968}. This model emphasizes the importance of reliability, which refers to the consistency or stability of measurement. Reliability is commonly assessed using indices such as Cronbach's alpha \citep{Cronbach1951}, which estimates internal consistency by evaluating the degree to which test items correlate with one another.

In parallel, Validity Theory addresses whether a test truly measures the construct it claims to assess. Central to this is the concept of latent constructs---unobservable traits or attributes such as intelligence, attitudes, or personality traits \citep{Furr2021}. Validity is not a single property but a collection of evidential approaches. Convergent validity assesses whether measures that should be related are indeed correlated, while discriminant validity ensures that measures of distinct constructs are not unduly correlated. Predictive validity evaluates how well scores forecast future performance, and external validity concerns the generalizability of results across populations and contexts.

To support construct validity, factor analysis is used to explore the underlying structure of item sets. Factor loadings indicate how strongly each item correlates with a latent factor; values above .60 are typically viewed as strong. Beyond Cronbach's alpha, Composite Reliability (CR) offers a more robust measure of internal consistency by accounting for individual item contributions, with acceptable values typically exceeding .70. Average Variance Extracted (AVE) measures the proportion of variance captured by the construct relative to error, with values of .50 or higher indicating adequate convergent validity. Together, these statistical tools and conceptual models form the backbone of rigorous measurement and evaluation of LLMs.

Cognitive Load Theory (CLT) is drawn upon to address stakeholders' demand for ease of use. CLT posits that learning is constrained by the limited capacity of working memory and emphasizes the importance of instructional design in managing cognitive load. CLT identifies three types of load---intrinsic, extraneous, and germane---that jointly determine the total load on working memory \citep{Sweller1988, Chandler1991}. Intrinsic load reflects the inherent complexity of the material, determined by the number and interactivity of elements that must be processed simultaneously. High element interactivity increases intrinsic load because each element interacts with---and thus depends on---others for meaning \citep{Sweller1988}. Extraneous load arises from the way information is presented and can be modified by instructional design \citep{Chandler1991}. Poorly designed materials---such as split-attention formats or redundant text---impose unnecessary processing demands that detract from learning. Germane load is the mental effort devoted to schema construction and automation, facilitating transfer to long-term memory \citep{Sweller1988, Paas2020}. Unlike extraneous load, germane load is considered beneficial and should be maximized once intrinsic and extraneous loads are managed.

\section{Design Cycle}

The Design Cycle translates the design objectives---shaped by the problem context and theoretical foundations from the Relevance and Rigor Cycles---into the concrete implementation of the PsyCogMetrics\texttrademark{} AI Lab, operationalizing them through the principles and processes of design science methodology.

\subsection{Design Objectives}

The Design Objectives and their accompanying Success Metrics (as shown in Table~\ref{tab:objectives}) emerge directly from the dual imperatives of the Relevance and Rigor cycles.

The Robust Evaluation objective and success metrics of addressing benchmark saturation, data contamination, and lack of coverage problems are derived from the Relevance Cycle's identification of three real-world pain points of LLM evaluation whose resolution directly serves AI developers' needs. Scientific Rigor is the design objective grounded in Popper's falsifiability and classical psychometric theory, ensuring every evaluation can be systematically scrutinized and potentially refuted. Accordingly, success is measured by reproducibility/repeatability/replicability---to confirm results technically, operationally, and independently---and by reliability and validity metrics (e.g., Cronbach's~$\alpha$, convergent/discriminant/predictive validity) drawn from Classical Test Theory and Validity Theory. Within the Relevance Cycle, the objective for Explainability emerges directly from stakeholder consultations---especially end users, regulators, and policymakers---who repeatedly cited a lack of transparent and interpretable model behavior as a barrier to trust and adoption. The design objective of Usability arises from the Relevance Cycle, where non-technical stakeholders need accessible, easy-to-use tools for LLM evaluation. The success metrics---minimize intrinsic load, minimize extraneous load, maximize germane load---are derived from the Rigor Cycle, specifically Cognitive Load Theory, which guides the platform's design to optimize cognitive efficiency and support effective user interaction. The Integration design objective and its Success Metrics---infrastructure/model, data, and evaluation/reporting integration---emerge from stakeholder needs identified in the Relevance Cycle, where fragmented tools limit usability, and from the Rigor Cycle, which demands reproducibility, psychometric validity, and cognitive accessibility. Together, these cycles highlight the necessity of a unified, scientifically grounded platform for evaluating LLMs.

\begin{table}[htbp]
\centering
\caption{Design Objectives and Success Metrics}
\label{tab:objectives}
\begin{tabular}{p{3.5cm} p{9cm}}
\toprule
\textbf{Design Objectives} & \textbf{Success Metrics} \\
\midrule
Robust Evaluation &
  Mitigate benchmark saturation problem \newline
  Mitigate data contamination \newline
  Mitigate coverage problem \\ \hline
Scientific Rigor &
  Reproducibility \newline
  Repeatability \newline
  Replicability \newline
  Reliability metrics \newline
  Validity metrics \\ \hline
Explainability &
  Transparency \newline
  Interpretability \\ \hline
Usability &
  Ease-of-Use \newline
  Minimize intrinsic load \newline
  Minimize extraneous load \newline
  Maximize germane load \\ \hline
Integration &
  Infrastructure \& model integration \newline
  Data integration \newline
  Evaluation, Metrics \& Reporting \\
\bottomrule
\end{tabular}
\end{table}

\subsection{Methodology}

Design-Science Research (DSR) aims to ``create what is effective'' by constructing purposeful information technology artefacts that address real-world problems \citep{Hevner2004}. Multiple process models operationalize these principles. \citeauthor{Peffers2006}'s \citeyearpar{Peffers2006} Design-Science Research Process (DSRP) prescribes sequential steps---problem identification, objective formulation, artefact design/development, demonstration, evaluation, and communication---while accommodating various entry points. Action Design Research (ADR) contends that building, intervention, and evaluation are inherently intertwined in organizational contexts, advocating for researcher participation as practitioners to rapidly iterate artefact designs \citep{Sein2011}.

The PsyCogMetrics\texttrademark{} AI Lab was constructed through nested Build--Intervene--Evaluate (BIE) loops that align with Action Design Research while remaining traceable to \citeauthor{Peffers2006}'s \citeyearpar{Peffers2006} Design-Science Research Process. The resulting platform is publicly available at \url{https://psycogmetrics.ai}, providing researchers direct access to its evaluation capabilities. To manage the multi-layered complexity of LLM evaluation, this study further orchestrates iterations across ``design echelons,'' a recent complexity-handling extension to DSR that decomposes large projects into coherent, hierarchically linked sub-artifacts \citep{Tuunanen2024}. Each BIE loop began with a minimum-viable slice of a target echelon and concluded with formative evaluation against design objectives and success metrics (Table~\ref{tab:objectives}). Lessons learned were immediately fed into the next loop, ensuring ``reflection-in-action'' and rigorous knowledge extraction \citep{Sein2011}. This cadence preserved DSR's build--evaluate symmetry \citep{Hevner2004} while allowing ADR's organizational interventions to surface emergent requirements \citep{Mullarkey2018}.

\subsection{Build}

\begin{figure}[htbp]
\centering
\includegraphics{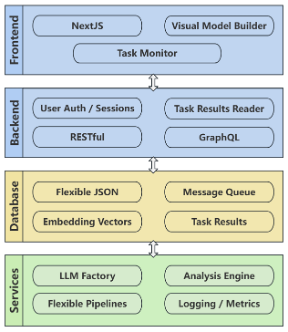}
\caption{System Architecture.}
\label{fig:architecture}
\end{figure}

To cope with the platform's complexity, we adopted the design-echelons methodology \citep{Tuunanen2024} and structured the system into four vertical design-echelons or layers, each serving a specific and essential role (Figure~\ref{fig:architecture}):

\textbf{Frontend Layer:} The Frontend layer provides an intuitive user interface and responsive interactions, building on modern Next.js framework with server-side rendering (SSR) to enhance performance and reduce load time. Interactive visual editing capabilities, such as draggable structural model building canvas, offer immediate feedback and intuitive operations, simplifying interactions and reducing cognitive load, compared to editing specifications in a domain-specific language. Real-time updates and streaming responses ensure dynamic responsiveness, crucial for managing complex experiments.

\textbf{Backend Layer:} The backend layer manages user authentication, authorization, sessions, project data and task tracking. It offers structured APIs exposing both RESTful and GraphQL endpoints to access user profile and project information, monitor task status, and interpret task results executed by the service layers. Serving as the gateway to the Database and Service layer, this loosely coupled integration avoids direct connection between UI and services, enhancing modularity and maintainability.

\textbf{Database Layer:} The Database layer acts as the central storage and communication hub, leveraging PostgreSQL's extensibilities to manage user information, project data, embedding vectors, task queues and analysis results. Its support of JSON data types enables flexible schemas, allowing addition or removal of non-essential fields compatible with new experiments, without compromising the overall data processing or workflows. Besides the traditional connection pooling and failover mechanisms, it offers robust support suited for AI applications and easy scalability.

\textbf{Service Layer:} The Service layer manages computationally intensive, long-running tasks asynchronously, separating complex workflows from interactive user interface. It communicates with the web application via the database instead of direct HTTP connections, ensuring service changes do not impact the web application, thus facilitating maximum flexibility, extensibility and scalability. This design allows customization of various experiments and integration of diverse large language models (LLMs), as long as task output remains consistent. The Service layer includes a customizable pipeline for simulation-based experiments, an LLM factory for interfacing with any LLM models, an analysis engine to run any necessary analysis, and comprehensive logging mechanisms to track usage and collect training data for future LLM optimization.

\subsection{Intervene}

In ADR's iterative Build--Intervene--Evaluate (BIE) loop, the Intervene phase is where the nascent IT artifact is introduced into its target organizational setting so that its real-world effects can be observed, and so that users and other stakeholders can engage with it directly. This phase serves two key purposes: (a)~to enact change in practice---testing the artifact's utility and uncovering unanticipated consequences---and (b)~to generate rich feedback that drives subsequent redesign. As \citet{Sein2011} note, the artifact is first deployed in a ``light-weight'' intervention (often using an alpha version in a limited context) and, once stabilized, moved into broader organizational settings as a beta version for more comprehensive evaluation and shaping.

This study adopts the self-as-user ``dog-fooding'' intervention strategy. This strategy has been discussed in the article ``Eating Your Own Dog Food'' by \citet{Harrison2006}, which traces the term's origins and shows how Microsoft and Eclipse have used the approach to test their own products internally before public release---serving as a controlled, lightweight intervention. Dogfooding is recognized as a powerful, low-latency feedback mechanism in product development \citep{Luke2017}. In the context of this study, it involves researchers on the team using the PsyCogMetrics\texttrademark{} AI Lab to carry out various LLM evaluation projects, thereby assessing the artifact's practical utility through firsthand experience.

We operationalize the ``dogfooding'' intervention by conducting an LLM evaluation study using the PsyCogMetrics\texttrademark{} AI Lab platform, which facilitates the elicitation, capture, and analysis of data from both artificial and human agents. A detailed account of the intervention is provided in a separate publication \citep{LiPsychometrics2025}.

In the study, we adapted key constructs from the Technology Acceptance Model (TAM)---Perceived Usefulness (PU), Perceived Ease of Use (EOU), and Purchase Intention (PI)---into a set of questions suitable for both LLM interrogation and human survey administration. Each item was presented as a standalone prompt, requesting a self-contained, numerical response on a 7-point Likert scale. For artificial agents, we selected two families of LLMs: OpenAI's GPT-3.5-turbo and GPT-4o, and Meta's LLaMA-2-13B-chat and LLaMA-3-8B-instruct, all accessed via the OpenRouter API (OpenRouter.ai). Under uniform experimental conditions, each model generated 500 responses across the full TAM item set using the diffusion mechanism described by \citet{LiPsychometrics2025}. In parallel, we recruited 248 human participants who met predefined criteria for purchase frequency and attention-check. Data quality was maintained through embedded veracity checks \citep{Coyle2001}.

Following the data collection, we conducted factor analysis and partial least squares (PLS)-based structural equation modeling, along with a series of reliability and validity checks. These interventions laid the groundwork for the subsequent phase of the BIE cycle, the Evaluate phase, during which the results were assessed in relation to the original design objectives and success metrics.

\subsection{Evaluate}

The Evaluate phase in the Build--Intervene--Evaluate (BIE) loop is concerned with assessing how well the newly built artifact works when deployed in its intended organizational setting. This evaluation was conducted using the publicly available PsyCogMetrics\texttrademark{} AI Lab (\url{https://psycogmetrics.ai}). Crucially, the design objectives and success metrics derived during the Relevance and Rigor cycle serve as the explicit criteria against which evaluation data are compared.

\textbf{Objective 1: Robust Evaluation.} Our AI Lab's novel evaluation framework effectively addresses three challenges in large language model assessment. First, it mitigates benchmark saturation by introducing psychometric measures that LLMs have not yet reached the ceiling score. As shown in Table~\ref{tab:validity}, for predictive validity, GPT-4o attains an $R^{2}$ of .443 for Purchase Intention, and LLaMA-3 attains .373 versus .599 for humans. For $R^{2}$, a measure on a scale between 0 and 1, this is a significant discrepancy. External validity is indicated by path coefficients of .46 (Perceived Usefulness $\rightarrow$ Purchase Intention) for GPT-4o and .46 for LLaMA-3, compared to .22 for human participants. For Ease of Use (EOU $\rightarrow$ Purchase Intention), GPT-4o scores .30 and LLaMA-3 scores .19---demonstrating a statistically significant difference from human participants, who score .65. Second, it overcomes data contamination concerns by employing reliability and construct-validity measures that focus on internal consistency within model outputs; because there is no single ``correct'' response to these questionnaires, any leakage of items into training data does not compromise the integrity of the assessment. Finally, it addresses the coverage problem by leveraging thousands of established psychological and cognitive assessment instruments originally designed for human testing, thereby ensuring comprehensive measurement of diverse cognitive and behavioral patterns when evaluating LLMs.

\textbf{Objective 2: Scientific Rigor.} The system ensures reproducibility by recording every step---from questionnaire design through data collection and analysis---as immutable, versioned events in a central database so the entire workflow can be replayed and shared to yield identical results. PsyCogMetrics\texttrademark{} AI Lab's reliability pipeline automatically computes Cronbach's~$\alpha$ to assess internal consistency, applies Shapiro--Wilk and Kolmogorov--Smirnov tests to each variable to check normality, and generates a composite data-quality score by combining missing-data rates, outlier prevalence, and model convergence diagnostics; for validity, it reports convergent validity, discriminant validity, predictive validity, external validity, Composite Reliability and Average Variance Extracted; it also reports full SEM fit indices (CFI, TLI, RMSEA, AIC, BIC), extracts and significance-tests all path coefficients against theoretical benchmarks, and executes a multi-model statistical validation framework---re-analyses under CFA, regression diagnostics, and comparisons to external benchmarks---to flag any discrepancies.

\textbf{Objective 3: Explainability.} PsyCogMetrics\texttrademark{} AI Lab ensures transparency by persistently logging every processing step, parameter, and callback in its event-sourced, database-centric architecture and by exposing live task statuses and detailed audit trails through its real-time UI. At the same time, it promotes interpretability by offering an intuitive, visual structural equation model editor in place of code. Alongside reliability and normality statistics, the system generates interpretations of the statistics such as the criteria of factor loadings ($\geq .50$; AVE $\geq .50$) and reliability thresholds (Cronbach's $\alpha \geq .70$; composite reliability $\geq .70$) to assist interpretation of the results. The AI Lab enables seamless export of datasets and model specifications in CSV and Lavaan syntax for external review and cross-validation.

\textbf{Objective 4: Usability.} PsyCogMetrics\texttrademark{} AI Lab achieves ease-of-use by offering a drag-and-drop SEM editor, end-to-end pipeline automation, and project-based organization so users never write boilerplate code or juggle tools; it minimizes intrinsic load by encapsulating structural equation model computations and psychometric tests behind a simple visualized interface; it cuts extraneous load through a database-centric, asynchronous task engine and a consistent UI component library that hides backend complexity; and it maximizes germane load by providing real-time fit statistics, customizable pipeline callbacks, and detailed audit trails that reinforce schema building through active, reflective engagement.

\textbf{Objective 5: Integration.} PsyCogMetrics\texttrademark{} AI Lab frees researchers from the hassle of provisioning and configuring software and hardware required for evaluating large language models. For infrastructure and model integration, it is built on a decoupled, microservice-driven architecture in which a Next.js frontend and RESTful APIs communicate exclusively through a central PostgreSQL database (via Prisma ORM), while Python workers poll for task entries and invoke a wide variety of LLMs available on the market (e.g., OpenAI GPT-4.5, Google Gemini 2.5 Pro, Anthropic Claude 3.7 ``Sonnet,'' xAI Grok-3, Meta Llama-4, etc.). For data integration, all questionnaire definitions, persona metadata, synthetic responses, and statistical outputs are normalized in the database and linked via an event-sourced audit log, ensuring complete end-to-end lineage. For evaluation, metrics, and reporting integration, the asynchronous Validation Engine automatically runs multi-model checks and writes metrics back for reporting.

Finally, real-time status updates and key psychometric metrics are exposed through a Task Monitor UI (via live API queries of the database), and all data and outputs can be exported in CSV or lavaan-compatible formats for further analysis or dashboarding, tightly weaving computation into the reporting workflow.

\begin{table}[htbp]
\centering
\caption{Predictive Validity and External Validity}
\label{tab:validity}
\begin{tabular}{lccccc}
\toprule
 & \textbf{GPT-3.5} & \textbf{GPT-4o} & \textbf{LLaMA-2} & \textbf{LLaMA-3} & \textbf{Human} \\
\midrule
PU $\rightarrow$ PI & 0.39*** & 0.46*** & 0.30*** & 0.46*** & 0.22*** \\
EOU $\rightarrow$ PI & 0.11* & 0.30*** & 0.21*** & 0.19*** & 0.65*** \\
$R^{2}$ of PI & 18.4\% & 44.3\% & 19.7\% & 37.3\% & 59.9\% \\
\bottomrule
\multicolumn{6}{l}{\footnotesize Note: * $P<0.05$; ** $P<0.01$; *** $P<0.001$}
\end{tabular}
\end{table}

\section{Discussion}

\citeauthor{Hevner2004}'s \citeyearpar{Hevner2004} seven-guideline framework clarifies that DSR requires (1)~an innovative artefact, (2)~clear problem relevance, (3)~rigorous evaluation, (4)~demonstrable novelty, (5)~formally represented design, (6)~a search process, and (7)~effective communication. This section discusses the contribution of this study as design science research.

(1)~PsyCogMetrics\texttrademark{} AI Lab is among the first integrated, cloud-based platforms that operationalize psychometric and cognitive-science theory for LLM evaluation. While several recent studies and toolkits introduced frameworks, benchmarks, or code libraries for probing LLM ``psychology,'' none offer end-to-end orchestration, event-sourcing, persona-injection, visual SEM editing, and mixed-method evaluation pipeline of PsyCogMetrics\texttrademark{} AI Lab. (2)~This study demonstrates problem relevance by identifying scientists' need for integrated, theory-driven tools. It mapped stakeholder needs---showing that AI developers require non-saturating, contamination-resistant benchmarks, regulators and end users demand transparency and robust safety metrics, and social scientists need minimal-code, accessible platforms---and systematically distilled these into concrete objectives and success metrics. (3)~By grounding each objective in Popperian falsifiability, Classical Test Theory, and Cognitive Load Theory during the Rigor cycles, the study commits to theory-driven, falsifiable success criteria, and then in the Evaluate phase empirically verifies them using psychometric reliability and validity. (4)~The study's novelty lies in its integration of psychometric adaptive testing and cognitive-science benchmarks to overcome ceiling effects and data-leakage issues existing LLM tools cannot address; (5)~its formally represented design is ensured by a layered system architecture; (6)~the search process is embodied in nested Build--Intervene--Evaluate loops across hierarchically organized design echelons, systematically exploring and refining features within an Action Design Research framework; (7)~and effective communication is achieved through a three-cycle narrative, with plans for tutorials, training programs, and workshops to introduce PsyCogMetrics\texttrademark{} AI Lab to the community.

\section{Conclusion}

In conclusion, this paper reports a design-science study that developed PsyCogMetrics\texttrademark{} AI Lab (\url{https://psycogmetrics.ai}), a rigorously designed, action-oriented artifact that addresses urgent and multifaceted challenges in LLM evaluation through a three-cycle Design Science Research methodology. By grounding development in stakeholder needs (Relevance Cycle), philosophical and psychometric theories (Rigor Cycle), and iterative build--intervene--evaluate loops (Design Cycle), the lab delivers a robust, reproducible, explainable, and user-friendly platform for evaluating LLMs from a cognitive science perspective. It moves beyond static benchmarks and opaque performance metrics to offer a transparent, scientifically valid framework that empowers AI developers, regulators, and cognitive researchers alike. Through its technical sophistication and theoretical grounding, PsyCogMetrics\texttrademark{} AI Lab not only advances the state of LLM evaluation but also establishes a replicable model for future design science interventions in AI research.

\section{References}

\begingroup
\renewcommand{\section}[2]{}  

\endgroup

\end{document}